\documentclass[a4paper]{jpconf}

\usepackage{graphicx}
\usepackage{citesort}

\begin{document}

\title{Gap-like feature in the normal state of $\bf{X(Fe_{1-x}Co_x)_2As_2}$, $\bf{X=Ba,Sr}$ and $\bf{Fe_{1+y}Te}$ revealed by Point Contact Spectroscopy}

\author{H. Z. Arham$^1$, C. R. Hunt$^1$, W. K. Park$^1$, J. Gillett$^2$, S. D. Das$^2$, S. E. Sebastian$^2$, Z. J. Xu$^3$, J. S. Wen$^3$, Z. W. Lin$^3$, Q. Li$^3$, G. Gu$^3$, A. Thaler$^4$, S. L. Budko$^4$, P. C. Canfield$^4$, L. H. Greene$^1$}

\address{\textsuperscript{$1$}Department of Physics and the Frederick Seitz Material Research Laboratory, University of Illinois at Urbana-Champaign, Urbana, Illinois 61801, USA 

\textsuperscript{$2$}Cavendish Laboratory, J. J. Thomson Ave, University of Cambridge, UK

\textsuperscript{$3$}Condensed Matter Physics and Materials Science Department, Brookhaven National Laboratory, Upton, New York, 11973, USA

\textsuperscript{$4$}Ames Laboratory and Department of Physics and Astronomy, Iowa State University, Ames, IA, 50011, USA}

\ead{arham1@illinois.edu}

\begin{abstract}
Point contact spectroscopy reveals a gap-like feature above the magnetic and structural transition temperatures for underdoped $Ba(Fe_{1-x}Co_x)_2As_2$, $SrFe_2As_2$  and $Fe_{1+y}Te$. The conductance spectrum starts showing an enhancement at temperatures as high as 177 K for $BaFe_2As_2$ ($T_N$ $\sim$ 132 K) and 250 K for $SrFe_2As_2$ ($T_N$ $\sim$ 192 K). Possible origins for this enhancement are discussed in light of recent experimental claims of nematicity in these materials. We construct a modified phase diagram for Co-doped Ba122 showing a gap-like feature existing above $T_N$ and $T_S$ for the underdoped regime.
\end{abstract}

\section{Introduction and Background}
The parent compounds of the iron pnictide and iron chalcogenide superconductors are multiband itinerant antiferromagnets ~\cite{Kamihara,Rotter}. A structural transition from tetragonal to orthorhombic or monoclinic symmetry slightly precedes or occurs simultaneously with the antiferromagnetic transition ~\cite{Cruz,Lester,Pratt,Bao}. The suppression of this antiferromagnetic ground state by various means \cite{Ren} leads to the emergence of superconductivity. As such, their phase diagram is strikingly similar to that of the copper based superconductors \cite{Basov}. For cuprate superconductors, a pseudo-gap phase precedes superconductivity, and opinions vary as to whether it is a forerunner or competitor to the superconducting state ~\cite{Timusk,Deutscher}. The debate for the existence of a similar pseudo-gap state in iron based superconductors is unresolved \cite{Johnston}. However, recent experimental work has produced evidence for nematicity in these materials in the form of electronic, spin and orbital anisotropy that does not correspond to the underlying crystal lattice symmetry and persists into the tetragonal paramagnetic state of these compounds ~\cite{Chuang,Chu,Tanatar,Yi,Harriger,Dusza}.

A point contact is simply a contact between two metals whose characteristic size $d$ is much less than the electron elastic and inelastic mean free paths: $l_{el},l_{in}>>d$. Point contact spectroscopy (PCS) studies the non-linearities of the current-voltage ($I-V$) characteristics of these metallic constrictions. If both metals are ohmic and uncorrelated, Harrison's theorem \cite{Harrison} dictates that $I-V$ curves for the junction shall be linear and $dI/dV$ shall be constant. Any nonlinearities in the conduction spectra are therefore a consequence of strong electronic correlations in the sample. PCS on a normal metal/superconductor junction shows a sub-gap conductance enhancement due to Andreev reflection \cite{Andreev}. This enhancement disappears at the superconducting transition temperature $T_c$. In recent years, PCS has also been used to probe heavy fermion compounds where it has been shown to be sensitive to the onset of the Kondo Lattice as a Fano lineshape, antiferromagnetic ordering, and the hybridization gap ~\cite{WKPsum}.

Thus far, PCS results reported on the iron-based compounds focus primarily on their superconducting phase ~\cite{Daghero,PCSsum}. In this study, we use PCS to probe the \emph{normal} state of the $\mathrm{Ba(Fe_{1-x}Co_x)_2As_2}$, $\mathrm{SrFe_2As_2}$ and $\mathrm{Fe_{1+y}Te}$. Our results reveal a gap-like feature (GLF) well above the magnetic ($T_N$) and structural ($T_S$) transition temperatures of these compounds. The onset temperature of this feature as reflected in the $dI/dV$ spectra allows us to identify a new region in the underdoped side of the $\mathrm{Ba(Fe_{1-x}Co_x)_2As_2}$ phase diagram. Superconductivity or long range magnetic order cannot be the source of our conductance enhancement. We relate our results to recent studies of electronic nematicity observed in these materials ~\cite{Chuang,Chu,Tanatar,Yi,Harriger,Dusza}. 
\section{Materials and Experimental Methods}

Single crystals of $\mathrm{Ba(Fe_{1-x}Co_x)_2As_2}$ are grown out of FeAs self flux as described in \cite{Sebastian} (for x = 0, 0.015, 0.025, 0.05, 0.055, 0.07 and 0.08) and \cite{Ni} (for x = 0.125). $\mathrm{Fe_{1+y}Te}$ single crystals are grown by a horizontal unidirectional solidification method. Metallic junctions are formed on freshly cleaved $c$-axis crystal surfaces and the differential conductance, $G(V)=dI/dV$, across each junction is measured using a standard four-probe lock-in technique. PCS is carried out in two different configurations: the needle-anvil method and the soft PCS method ~\cite{Naidyuk,Daghero}. In the needle-anvil setup, an electrochemically polished Au tip is brought into gentle contact with the sample. For soft PCS, we sputter 50-100{\AA} $\mathrm{AlO_x}$ on our crystals to act as an insulating barrier. Using Ag paint as a counter electrode, parallel, nanoscale channels are introduced for ballistic current flow by fritting \cite{Holm} across the oxide layer. Similar spectra are obtained from both PCS methods. Unlike needle-anvil PCS, soft PCS junctions are stable over wide temperature ranges. 

\section{Results}

\begin{figure}[thbp]
		\includegraphics[scale=0.795]{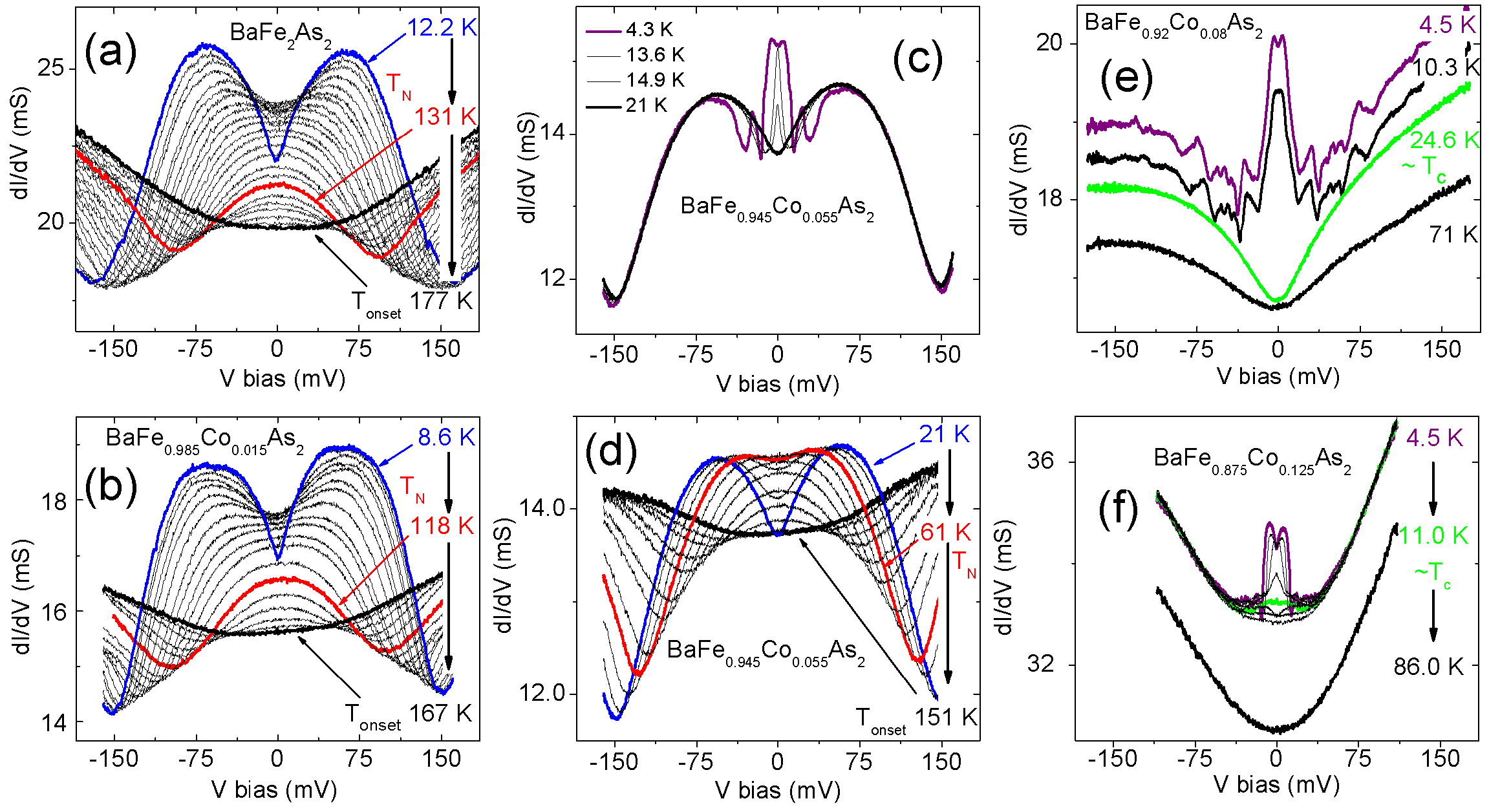}
	\caption{Conductance spectra for $\mathrm{Ba(Fe_{1-x}Co_x)_2As_2}$: All dI/dV curves display raw data. For (a), (b), and (d) the curves are taken at intervals of 10-15 K. The first column (a, b) shows nonsuperconducting underdoped compounds (x = 0, 0.015). Both crystals show a gap-like feature with a dip at zero bias and conductance peaks at $\sim$65 mV. The conductance enhancement is superimposed on a parabolic background and survives well above $T_N$ (red curves). The second column (c, d) is for x = 0.055, which exhibits coexistence of superconductivity and magnetism at low temperatures. At 4.3 K, conductance peaks due to Andreev reflection coexist with the high bias gap like feature. The third column (e, f) shows the spectra for the overdoped samples x = 0.08, 0.125. Low bias Andreev peaks are present at 4.5 K, and no conductance enhancement is observed above $T_c$.}
\end{figure}  

Figure 1 displays the temperature evolution of the conductance spectra for $\mathrm{Ba(Fe_{1-x}Co_x)_2As_2}$.  Figure 1a shows the spectra for undoped Ba122 ($T_N$ $\sim$ 132 K). At the lowest temperature (blue curve), we see a dip in $G(V)$ at zero bias and two asymmetric conductance peaks at $\sim$ 65 mV. This double peak GLF is superimposed on a parabolic background. As the temperature is increased, the dip at zero bias fills, the conductance peaks move inward, and the bias voltage range of the conductance enhancement decreases. No change in the conductance spectra is observed as $T_N$ is crossed (red curve), indicating that these features are not due to the long range magnetic order. The conductance enhancement eventually disappears leaving behind the parabolic background at 177 K, more that 40 K above $T_N$. Similar spectra are obtained from two other underdoped nonsuperconducting samples: x = 0.015 ($T_N$ = 117 K, Figure 1b); and x = 0.025 ($T_N$ = 103 K). The onset temperature for the conductance enhancement is slightly reduced as the Co doping is increased, reaching a value of $\sim$ 160 K for x = 0.025.  

Figures 1c and d show the conductance spectra for x = 0.055, where long-rage magnetic order exists above the superconducting dome ($T_c$ = 11.5 K, $T_N$ = 63 K). At the lowest temperature, the lower bias voltages ($<$ 15 mV) are dominated by Andreev reflection, with Andreev conductance peaks centered around $\sim$ $\pm$ 5mV. However, just like the parent compound, two conductance peaks occur at $\sim$ 60 mV. As the temperature is increased, Andreev reflection dies out and the high bias GLF evolves just like it does for the undoped parent compound. A second coexisting sample with x = 0.05 shows similar spectra ($T_c$ = 8.9 K, $T_N$ = 70 K). 

Figures 1e and f show the conductance spectra for overdoped samples with x = 0.08 ($T_c$ = 24 K) and x = 0.125 ($T_c$ = 11 K). At 4.5 K, conductance enhancement with split Andreev peaks is observed due to superconductivity. As the temperature is increased, the peaks move together and the Andreev signal decreases. All conductance enhancement disappears as $T_c$ is crossed and only a V-shaped background remains. Nothing akin to the high bias GLF seen for the underdoped samples is observed. The bumps at 50 mV and 75 mV at 4.5 K for 8$\%$ Co and the slight dip in the zero bias conductance immediately above $T_c$ for 12.5$\%$ Co are not reproducible and are possibly contact geometry effects \cite{Daghero}. A sample with x = 0.07 ($T_c$ = 22 K) also only shows Andreev reflection without any high bias gap like feature.

\begin{figure}[thbp]
		\includegraphics[scale=0.795]{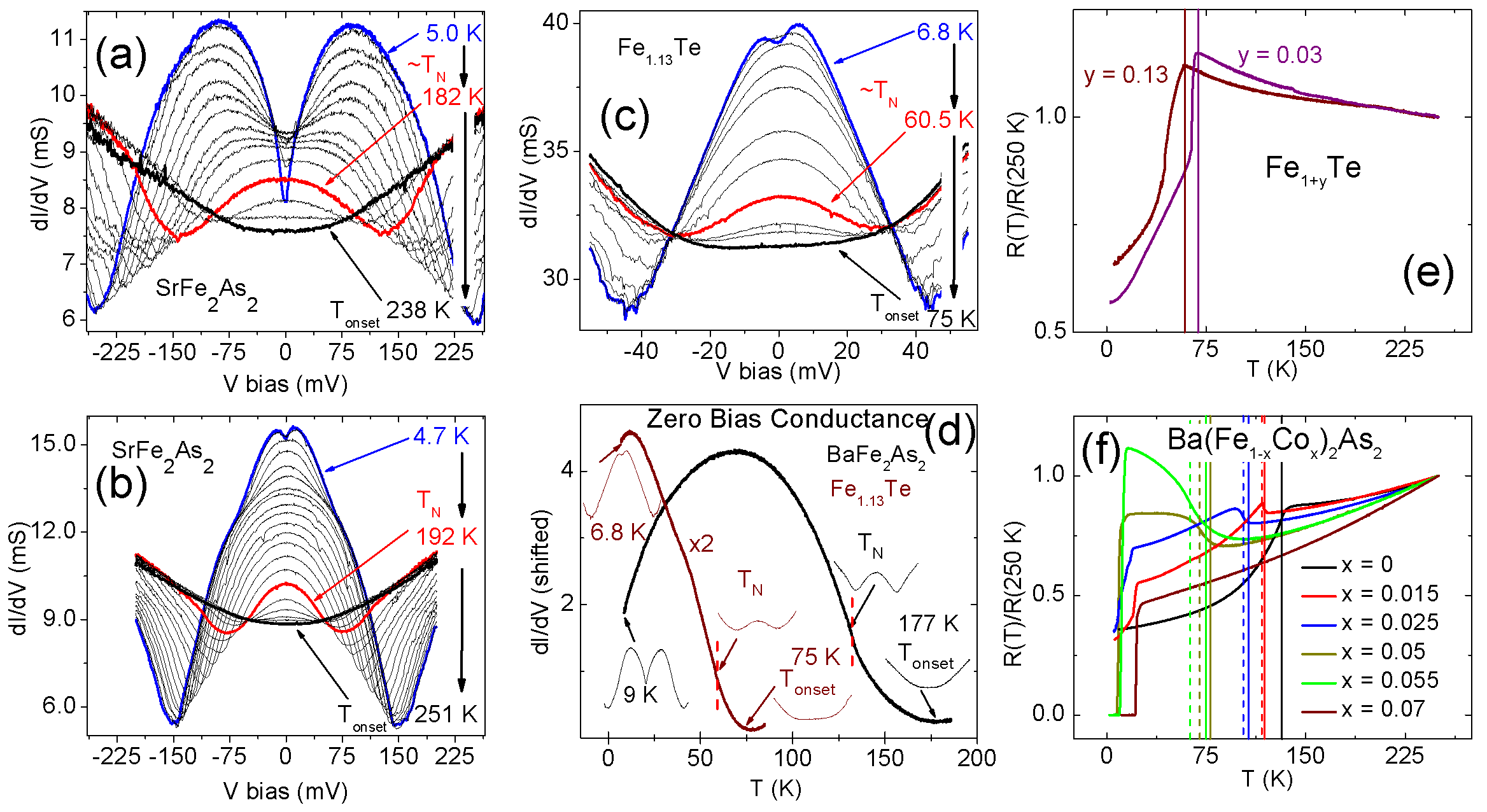}
	\caption{(a, b) The conductance spectra for $\mathrm{SrFe_2As_2}$ is similar to that of underdoped Ba122. The location for the conduction peak of the gap like feature varies across different junctions. However, the conductance enhancement always exceeds $T_N$ and its onset temperature is consistent across junctions. (c) $\mathrm{Fe_{1.13}Te}$  shows a conductance spectra similar to that of the iron pnictides. However the energy range for the conductance enhancement is smaller than that of Ba122 or Sr122. (d) The zero bias conductance (ZBC) vs. temperature also shows that the conductance enhancement lasts well above $T_N$. The insets correlate the spectra obtained at different temperatures to the ZBC curve. (e, f) Resistance vs. Temperature for $\mathrm{Ba(Fe_{1-x}Co_x)_2As_2}$ and $\mathrm{Fe_{1+y}Te}$. Upon cooling down from room temperature, resistance for Ba122 decreases while that for Fe11 increases. The solid lines represent $T_S$ and the dotted lines $T_N$. For Fe11, $T_S=T_N$.} 
\end{figure}  

The conductance peak position for the GLF varies from contact to contact. An extreme case of this is shown in Figures 2a and b. $\mathrm{SrFe_2As_2}$ has a $T_N$ of $\sim$ 192 K and its conductance spectrum is similar to that of underdoped Ba122. In Figure 2a, the conductance peak for the GLF occurs at $\sim$ 75mV at low temperature and the conductance enhancement remains until $\sim$ 240 K. However, in Figure 2b, the conductance peak occurs at $\sim$ 15 mV while there is a prominent hump in the spectra at $\sim$ 75 mV. The conductance enhancement for this junction persists to $\sim$ 250 K. Despite the wide variation in peak location, the onset temperature for the conductance enhancement is consistent to within $\sim \pm$15 K across different junctions for both Ba122 and Sr122. 

Similar conductance features are also observed in the parent compounds of the iron chalcogenides, $\mathrm{Fe_{1.13}Te}$ (Figure 2c) and $\mathrm{Fe_{1.03}Te}$ (not shown). While the conductance shape is similar, the energy scale is smaller. The range of conductance enhancement is $\sim$40 mV for $\mathrm{Fe_{1.13}Te}$. For the underdoped pnictides, the range of conductance enhancement is always larger than $\sim$ 125 mV.

The iron pnictides and the iron chalcogenides show very different temperature dependencies in resistivity and may be classified as bad metals \cite{Johnston}, making ballistic PCS challenging to carry out. Our resistance vs. temperature data are plotted in Figures 2e and f. Cooling down from room temperature, the resistance for $\mathrm{Fe_{1+y} Te}$ increases while that for $\mathrm{Ba(Fe_{1-x}Co_x)_2As_2}$ decreases. Both materials show a kink with a gradient change on crossing $T_N$ and $T_S$, following which the resistance for Fe11 also decreases, but at a much faster rate than that of Ba122. Despite differences in their R(T), the two families show similarly shaped conductance spectra. This indicates that our observed conductance feature cannot be accounted for by local heating effect, and our junctions are either in the ballistic or diffusive regimes ~\cite{Naidyuk,Daghero}. In the thermal regime, the junction resistance depends on the bulk resistivity of the crystals, and different shaped spectra for Ba122 and Fe11 would be seen. 

Figure 2d shows the conductance value at zero bias for $\mathrm{BaFe_2As_2}$ and $\mathrm{Fe_{1.13}Te}$. Again, it clearly shows that the enhancement lasts well above $T_N$, marked by the vertical red lines in the figure. The insets in the figure correlate the spectra obtained at different temperatures to the zero bias conductance curve. We define $T_{onset}$ as the temperature at which a conductance enhancement, superimposed on the parabolic background, is observed. Averaged over multiple contacts, $T_{onset}$ gradually decreases with increasing in Co concentration in $\mathrm{Ba(Fe_{1-x}Co_x)_2As_2}$. For $\mathrm{Fe_{1+y}Te}$, $T_{onset}$ decreases with increasing excess Fe concentration.

From our observations, together with $T_S$, $T_N$ and $T_c$ determined via our resistivity measurements, we construct a modified phase diagram for $Ba(Fe_{1-x}Co_x)_2As_2$ (Figure 3). The conductance enhancement in the normal state is only observed for the underdoped samples. The onset temperature is reduced as the Co doping is increased and it appears that this GLF ends at the same doping at which the magnetic and structure transitions disappear, suggesting the conductance enhancement is correlated with these transitions.

\begin{figure}[h]
\includegraphics[scale=0.2]{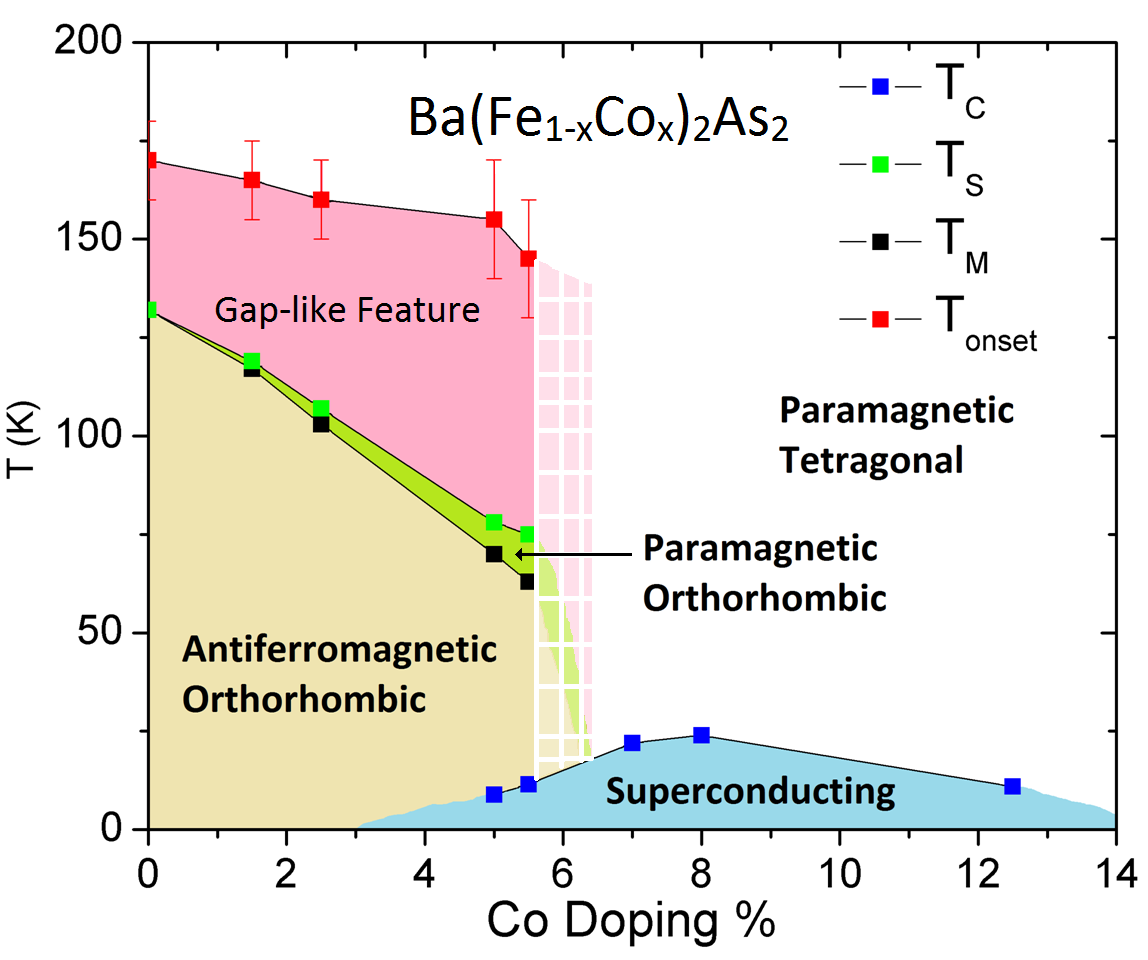}\hspace{2pc}%
\begin{minipage}[b]{14pc}\caption{\label{label}Modified phase diagram for $Ba(Fe_{1-x}Co_{x})_2As_2$.   The GLF region is determined by the onset of the PCS conductance enhancement and the other points are taken from our resistivity vs. temperature measurements.}
\end{minipage}
\end{figure}

\section{Discussion}
In our experiments, we probe twinned samples along the c-axis. The same signal with the same temperature evolution is observed for the pnictides and the chalcogenides which have different resistivity behavior and magnetic ordering wave vectors: ($\pi$,0) vs. ($\pi$,$\pi$). The different magnetic order emerges due to the different anion height for the pnictides and the chalcogenides \cite{Yin}. No definite consensus has been achieved as to what drives the structural transition in these materials. Electron orbital ordering ~\cite{Kruger,Chen3,Lv3,Lee} and fluctuating antiferromagnetism ~\cite{Fang,Yildirim} have been proposed as candidates.

Theoretical models predict a conductance enhancement for a normal metal/itinerant antiferromagnet junction due to a spin-dependent Q-reflection ~\cite{AndersenBobkova}. Since our enhancement survives above $T_N$, these models do not apply to it. Ineleastic neutron scattering reveals high energy ($>$100meV) spin excitations in the paramagnetic, tetragonal phase of twinned $\mathrm{BaFe_2As_2}$ suggesting a nematic state \cite{Harriger}. However, we do not know of a mechanism where spin fluctuations or nematicity can cause such a conductance enhancement. 

Evidence for normal state nematicity from detwinned samples is complicated by the symmetry breaking pressure applied to detwin the crystal. Normal state in-plane resistive anisotropy is only observed for detwinned underdoped pnictides ~\cite{Chu,Tanatar}. ARPES \cite{Yi} detects orbtial ordering, and optical conductivity \cite{Dusza} an in-plane anisotropy, both lasting well above $T_N$ for detwinned, underdoped Ba122. We observe a conductance enhancement for \emph{twinned} underdoped pnictides that lasts well above $T_N$ similar to these papers. This leads us to speculate about a similar micrscopic phenomenon, (possibly $d$-orbital ordering as seen by ARPES) driving our conductance enhancement. 

Scanning tunneling microscopy on a single domain of $\mathrm{Ca(Fe_{1-x}Co_x)_2As_2}$ \cite{Chuang} detects strong correlations at 4.3 K in the form of static unidirectional electronic nanostructures of dimension 8 times the inter-iron-atom distance along the a-axis. It would be interesting to know the temperature at which these electronic nanostructures set in and whether they occur for overdoped compounds or not. As Harrison's theorem \cite{Harrison} is not applicable to strongly correlated materials, it will be interesting to calculate $dI/dV$ curves based on the density of states and Fermi velocity of the orbitally or magnetically ordered state in the iron based compounds for comparison with our conductance spectra. 

In conclusion, we have performed PCS measurements on $\mathrm{Ba(Fe_{1-x}Co_{x})_2As_2}$, $\mathrm{SrFe_2As_2}$ and $\mathrm{Fe_{1+y}Te}$, and we observe a highly reproducible conductance enhancement that survives well above the magnetic and structural transition temperatures of these materials. The origin of this enhancement is not clear but we suggest it is related to the observations of electronic orthorombicity or orbital ordering observed in detwinned samples or the nematicity reported at high and low temperature in twinned crystals. Our measurements provide further experimental evidence that the normal state of the underdoped pnictides exhibits strong electronic correlations.  

This is work supported by the Center for Emergent Superconductivity, an Energy Frontier Research Center funded by the US DOE Award No. DE-AC0298CH1088. The work at BNL is carried out under U.S. DOE Award No. DE-AC0298CH10886. University of Cambridge is supported by EPSRC, Trinity College, the Royal Society and the Commonwealth Trust. Ames Lab is operated by ISU under DOE Contract No. DE-AC02-07CH11358.

\section*{References}
\bibliography{myrefs}

\end{document}